# GROUND-BASED LIGHTCURVE OBSERVATION CAMPAIGN OF (25143) ITOKAWA BETWEEN 2001 AND 2004

S. Nishihara [1,2], K. Kitazato[1,2], M. Abe[1], S. Hasegawa[1],
M. Ishiguro[3], H. Nonaka[1,2], Y. Ohba[4], T. Ozawa[5],
Y. Sarugaku[1,2], Y. Yadoumaru[5], C. Yoshizumi[6], M. Okyudo[7]

[1]*Institute of Space and Astronautical Science, Japan Aerospace Exploration Agency, 3-1-1 Yoshinodai, Sagamihara, Kanagawa 229-8501, Japan*
[2]*Graduate School of Science, University of Tokyo, 7-3-1 Hongo, Bunkyo-ku, Tokyo 113-0033, Japan*
[3]*Institute for Astronomy, University of Hawaii, 2680 Woodlawn Drive, Honolulu, Hawaii 96822, USA*
[4]*Accenture Japan Ltd., 7-1-16 Akasaka, Minato-ku, Tokyo 107-8672, Japan*
[5]*Misato Observatory, 180 Matsugamine, Misato-cho, Wakayama 640-1366, Japan*
[6]*Tokushima Science Museum, 45-22 Kibigatani, Itano-cho, Tokushima 779-0111, Japan*
[7]*Student Center for Independent Research in the Sciences, Wakayama University, 930 Sakaedani, Wakayama, Wakayama 640-8510, Japan*

**Abstract.** The asteroid (25143) Itokawa is a target object of the Japanese sample return mission, HAYABUSA. We have observed Itokawa in optical wavelength (R-band) with the 1.05-m Schmidt telescope at the Kiso Observatory, the 2.24-m telescope of University of Hawaii, and the 1.05-m telescope at the Misato Observatory since 2001. From the analysis of the data, we present the relationship between brightness and the solar phase angle, 6.9 to 87.8 deg. We obtained the absolute magnitude $H_R(0)$ = 19.09±0.37, and the slope parameter $G_R$ = 0.25 ± 0.29. The rotational period of Itokawa is 12.1324 ± 0.0001 hours.

## 1. Introduction

After failing to meet the launch window for (10302) 1989ML, the Japanese sample return mission, MUSES-C (HAYABUSA) was redesigned to rendezvous with an Apollo-type near-Earth object, (25143) Itokawa in 2000. The MUSES-C spacecraft, HAYABUSA, was successfully launched on May 9, 2003.

Its target, (25143) Itokawa made close approaches to the Earth in 2001 and 2004, during which we collected an extensive ground-based database of photometry. From optical photometry and radar observations during the 2001 apparition, the physical properties of this asteroid, such as the size, shape, and spin state were obtained. Dermawan et al. (2002) reported that its synodic rotational period was 12.13 ± 0.02 hours and that its visual absolute magnitude (H) and the slope parameter (G) were 18.61 ± 0.18 and 0.29 ± 0.14, respectively. Ohba et al.(2003) determined a pole orientation of $\lambda$ = 320° ±30° and $\beta$ = −75° ±12° and a triaxial ellipsoid shape with aspect ratios $a : b : c$ = 1 : 0.45(±0.09) : 0.26(±0.06), by assembling lightcurve data obtained with three telescopes at Kiso, Mitaka, and Pic du Midi. Kaasalainen et al. (2003) also compiled 53 lightcurves, then established a 3D shape model to find its rotational period was 12.132 ± 0.0005 hours.

In this paper, we present the results of photometric observations of Itokawa obtained by our Japanese observation team between 2001 and 2004. Photometric observations and data reduction are described in section 2. In section 3, we present the results and a discussion, including the rotational period and phase curve. We provide a summary in section 4.

## 2. Observations and Data Reduction

We carried out optical photometric observations of (25143) Itokawa during its 2001 and 2004 apparitions. During the 2001 apparition, we used the 1.05-m Kiso Schmidt telescope with the SITe 2kCCD (Abe et al. 2002 ; Ohba et al. 2003). On the other hand, during the recent apparition in 2004, we performed optical observations at the University of Hawaii 2.24-m telescope with a 8k mosaic CCD, at the Misato Observatory 1.05-m Cassegrainian telescope with ST-9 CCD, and also at the Kiso Observatory. The 2kCCD (2048 × 2048 pixels) of the Kiso Schmidt telescope was used to cover a field of view 51'×51'. In the observations with the 8k mosaic CCD (8192 × 8192 pixels) of the UH 2.24-m telescope, its field of view was 30' × 30', and we attempted 2 × 2 binning for all nights. The ST-9 CCD (512×512 pixels) of the 1.05-m Misato telescope gave a field of view about 9' × 9'. The observational conditions are summarized in Table 1. We used exposure times ranging from 60 *sec.* to 900 *sec.* for each raw image. As all the images of the asteroid obtained at the Misato Observatory were faint, we shifted and added the 20 images to increase the signal/noise ratio. An effective exposure time of the composite frame was 3600 *sec.*. Because the rotational period of this asteroid is close to 12 hours, its general trend in a part of the lightcurve will become apparent.

On every photometric night, G-type standard stars, listed in the Landolt (1992) catalog, were recorded more than twice a night. The other days, some of the background stars were used for comparison stars. Since we had at least one photometric night for each observational sequence, we could determined the reduced magnitudes of the asteroid obtained all observation nights using the standard stars and comparison stars. The raw image data were bias- or darksubtracted and normalized using flat-field images. We used the software *IRAF* for aperture photometry (*APPHOT*) to measure the entire image data, including the asteroid, standard stars and comparison stars.

All of the observed asteroid magnitudes $M(\alpha)$, at the phase angle $\alpha$, are then normalized to the reduced magnitudes $H(\alpha)$, by eliminating the distance dependence, as follows:

$$H(\alpha) = M(\alpha) − 5\log(r\Delta), \qquad (1)$$

Table 1. Observational circumstances and ephemeris data (J2000.0) for (25143)Itokawa.

| (UT) | $r$ (AU) | $\Delta$ (AU) | $\alpha$ ( ) | EcLon (°) | EcLat (°) | Site | Filter |
|---|---|---|---|---|---|---|---|
| 2001Mar26 | 1.018 | 0.040 | 59.4 | 187.1 | 1.5 | K | B,V,R,I |
| 2001Mar29 | 1.009 | 0.038 | 78.4 | 190.3 | 1.5 | K | B,V,R,I |
| 2001Mar31 | 1.003 | 0.039 | 83.2 | 192.5 | 1.5 | K | B,V,R,I |
| 2001Apr01 | 1.000 | 0.040 | 87.6 | 193.6 | 1.5 | K | B,V,R,I |
| 2001Aug22 | 1.305 | 0.319 | 20.3 | 335.5 | -1.7 | K | R |
| 2001Aug23 | 1.309 | 0.321 | 19.1 | 336.1 | -1.7 | K | R |
| 2001Aug24 | 1.313 | 0.323 | 18.0 | 337.4 | -1.7 | K | R |
| 2001Aug25 | 1.317 | 0.325 | 16.8 | 337.4 | -1.7 | K | R |
| 2003Dec01 | 1.528 | 0.786 | 34.5 | 95.7 | 0.7 | K | R |
| 2003Dec02 | 1.525 | 0.776 | 34.3 | 96.1 | 0.7 | K | R |
| 2003Dec03 | 1.522 | 0.765 | 34.0 | 96.6 | 0.7 | K | R |
| 2003Dec04 | 1.519 | 0.755 | 33.8 | 97.1 | 0.8 | K | R |
| 2004Jan15 | 1.369 | 0.396 | 11.3 | 119.0 | 1.3 | K | R |
| 2004Jan19 | 1.353 | 0.373 | 7.8 | 121.3 | 1.3 | K | R |
| 2004Jan20 | 1.348 | 0.368 | 6.9 | 121.9 | 1.3 | K | R |
| 2004Apr10 | 1.019 | 0.250 | 79.1 | 186.8 | 1.6 | K | R |
| 2004Apr11 | 1.016 | 0.248 | 79.9 | 187.9 | 1.5 | K | R |
| 2004Apt12 | 1.012 | 0.246 | 80.7 | 188.9 | 1.5 | K | R |
| 2004Sep06 | 1.304 | 0.348 | 27.3 | 335.2 | -1.6 | H | R |
| 2004Sep07 | 1.309 | 0.356 | 27.7 | 335.8 | -1.6 | H | R |
| 2004Sep08 | 1.313 | 0.363 | 28.1 | 336.5 | -1.6 | H | R |
| 2004Sep09 | 1.317 | 0.371 | 28.5 | 337.1 | -1.6 | H | R |
| 2004Sep12 | 1.330 | 0.393 | 29.6 | 339.0 | -1.6 | M | R |

$r$: Heliocentric distance (AU).
$\Delta$: Geocentric distance (AU). $\alpha$: Phase (Sun-Asteroid-Observer) angle (°).
EcLon: Observer-centered ecliptic longitude of the asteroid (°).
EcLat: Observer-centered ecliptic latitude of the asteroid (°).
Site: Observatory (K: Tokyo-Kiso, H: UH88-Mauna Kea, M: Misato-Wakayama).

where $r$ is the heliocentric distance of an asteroid, and $\Delta$ is the geocentric distance. Thus, the reduced magnitudes of the asteroid are now only connected with their phase angles ($\alpha$).

## 3. Results and Discussion

Since our data, which have been archived since the 2001 apparition, were collected under various circumstances, Figure 1 shows some composed lightcurves for each observational sequence. The lightcurve data for 2001 March (top left) have significantly varying mean magnitudes, amplitudes, and minimum times with respect to solar phase angles. Therefore no curve fitting was attempted. The lightcurve for 2003 December (middle left) is also not fitted, because the plotted data have relatively large error bars and only include a part of the rotational phase. The curve fit for 2004 April (lower left) displays an anomalous feature, which might result from its large solar phase angles.

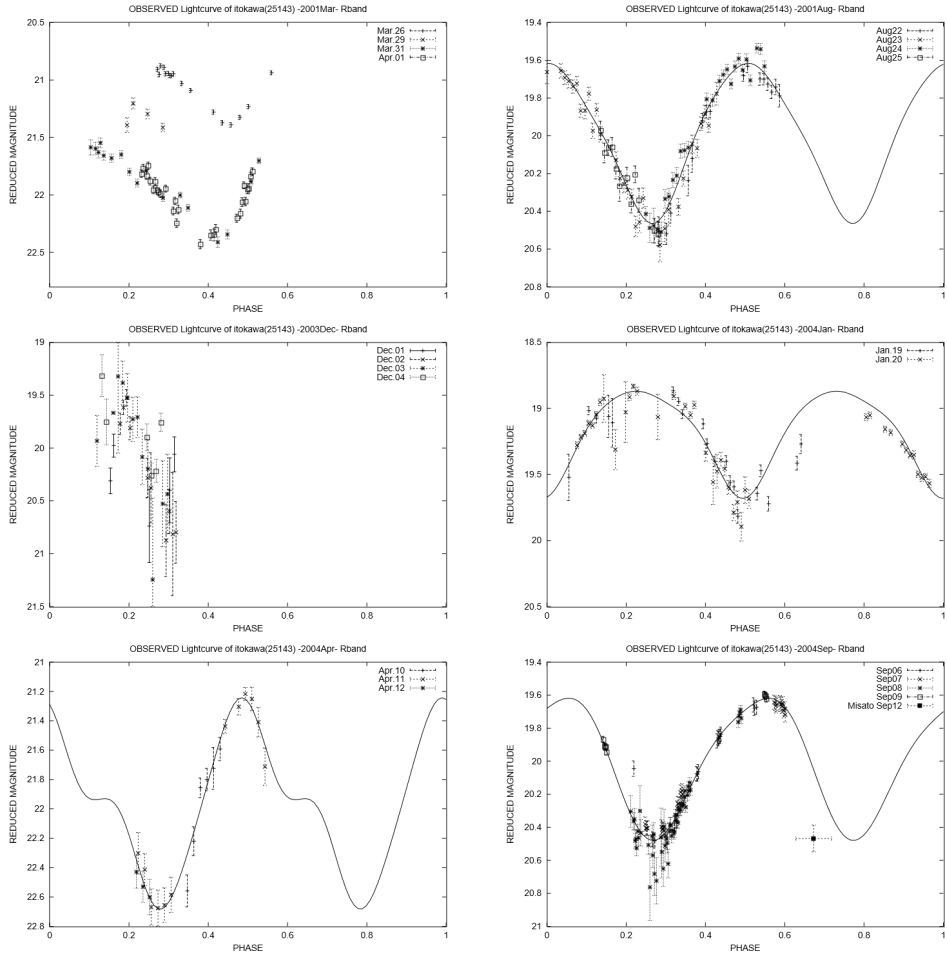

Figure 1. Composite lightcurves of (25143) Itokawa. The each observed data points for 2001 March 26 - April 01 (top left), 2001 August 22 - 25 (top right), 2003 December 01 - 04 (middle left), 2004 January 19 - 20 (middle right), 2004 April 10 - 12 (lower left), and 2004 September 06 - 12 (lower right), respectively, display with error bars. The rotational period is assumed 12.132 hours (Kaasalainen et al. 2003) with epochs set as exposure-starttimes of each first image of sequence. The horizontal axes show the rotational phase, the vertical axes give brightness in units of reduced magnitude, and its solid lines are sixth-order fitting obtained from the Fourier analyses.

### 3.1. Period Analyses

To obtain the rotational period at a certain level of precision, it is necessary to know which side of the body is being observed; so an ephemeris for physical observations is indispensable. The scattered light from the surface of a spherical object follows the reciprocity principle, and depends upon the geometry of a light source, an object and an observer. The "Standard Feature", (e.g. lightcurve maximum), varies with changes in the phase angle bisector (PAB), namely the bisector of the direction toward the observer and the Sun. To separate the sidereal rotational period from the apparent synodic rotational period, we need to assume obliquity of the spin axis of the asteroid. In this analysis we assumed the spin axis was perpendicular to the ecliptic plane and determined the astrocentric longitude of PAB considering results obtained by Ohba et al. (2003) and Kaasalainen et al. (2003). The light-travel times were also taken into account. In Figure 2, there are two interpolations of the lightcurve from the epoch 2001 August to 2004 September. To maintain consistency of the reduced magnitudes and amplitudes, we only show the observed lightcurves having similar value of $\alpha$. Fig.2(a) shows composite lightcurve assumed 12.132 hours, and Fig.2(b) assumed that the rotation period was 12.1324 hours. It is apparent that Fig.2(b) is the best fit result.

By modulating the rotational period for phase over 3 years, we derived 12.1324± 0.0001 hours. Our assumption that the obliquity of the spin axis was perpendicular to the ecliptic plane made it impossible to obtain the rotational period to four decimal places for each apparition. Nevertheless our obtained rotational period is more accurate than the result derived only from the observational dataset during the 2001 apparition; 12.132 ± 0.0005 hours (Kaasalainen et al., 2003).

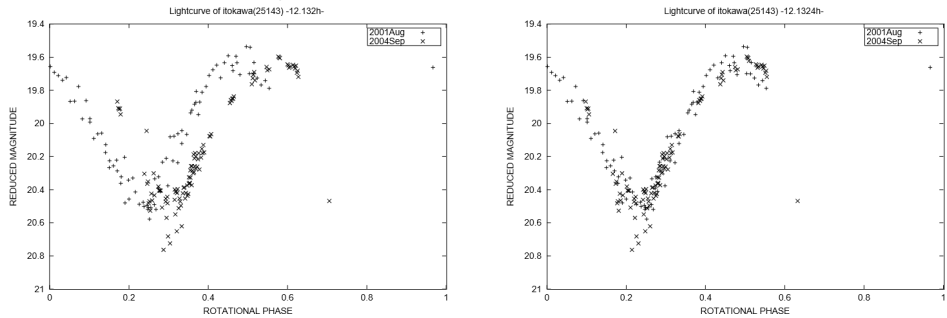

Figure 2. Compared rotational period of (25143) Itokawa. Left panel (a) shows composite lightcurve assumed 12.132 hours, and right panel (b) displays 12.1324 hours.

### 3.2. Phase Curve

The *H,G* magnitude system was adopted by IAU Commission 20 in 1985 (Marsden 1986) to describe the magnitude system for asteroids (Bowell et al., 1989). The mean magnitude of an asteroid can be calculated from the formula

$$H(\alpha) = H(0) - 2.5\log[(1 - G)\Phi_1(\alpha) + G\Phi_2(\alpha)], \qquad (2)$$

where $H(\alpha)$ is the reduced magnitude at solar phase angle $\alpha$. $H(0)$ is the *absolute magnitude*; that is, the reduced magnitude explicitly at $\alpha = 0°$. Commonly the mean V-band brightness is used in this method. $G$ is termed the *slope parameter*; which is indicative of the gradient of the phase curve. $\Phi_1$ and $\Phi_2$ are two specified phase functions, normalized to unity at $\alpha = 0°$; full descriptions of these phase functions are seen in Bowell et al. (1989).

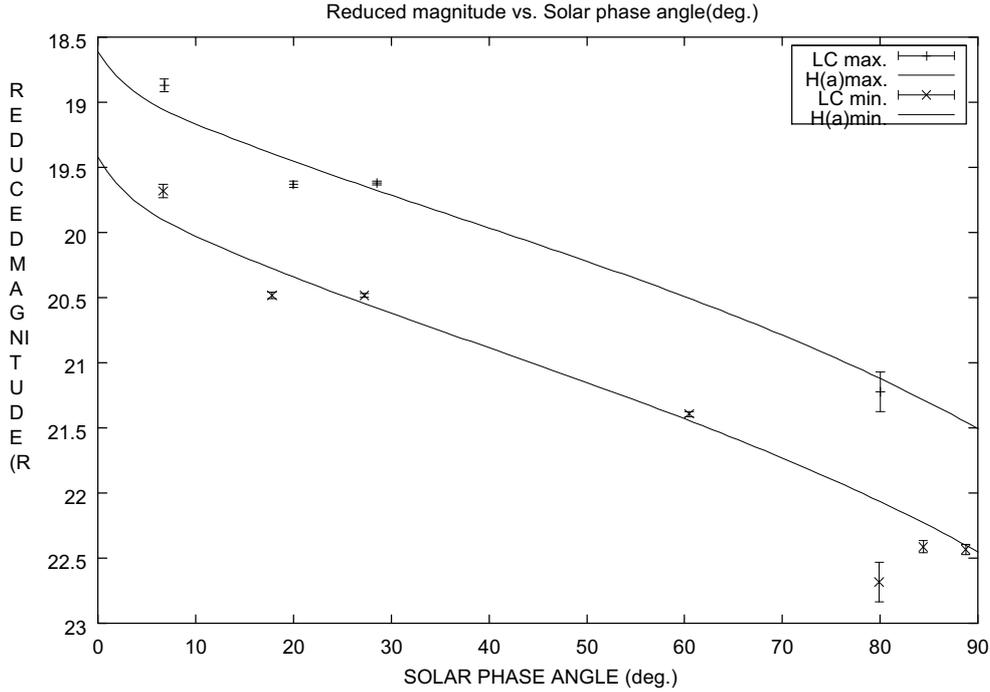

Figure 3. Phase curve of Itokawa. The lightcurve maximum and minimum of each observed sequence, only if the lightcurves had corresponding values, are plotted with error bars, respectively. The dashed lines are separately fitted by the least-squares analyses.

A phase curve of this asteroid is shown in Figure 3. We thus obtained $H_R(0)$ = 19.09±0.37, $G_R$ = 0.25±0.29 in R-band. Here, $H_R(0)$ was taken as the average of $H_R(0)max$ = 18.61 ± 0.48 and $H_R(0)min$ = 19.42 ± 0.21, and $G_R$ was the average of $G_R max$ = 0.29±0.39 and $G_R min$ = 0.21±0.13. From the visible and near-infrared spectroscopic measurements of (25143) Itokawa carried by Binzel et al. (2001), this asteroid was classified as S(IV)-type. Harris and Young (1988) and Tedesco (1989) showed that *G*-values for asteroids of C-type, S-type, and E-type lie in the ranges 0.15-0.21, 0.23-0.34, and 0.40-0.50, respectively. Thus, our result is in agreement with the detailed spectroscopic taxonomic results. Dermawan et al. (2002) reported $H_V$ = 18.61 ± 0.18, $G_V$ = 0.29 ± 0.14. Abe et al. (2002) also estimated absolute magnitude in V-band, $H_V$ = 19.73 ± 0.17, and slope parameter $G_V$ = 0.21±0.10. While the obtained absolute magnitude varies depending on the wavelength band and the rotational aspect angle of the asteroid, the *G* parameter is comparable. Considering this fact, the previous results of *G* values are consistent with our result. However, our observational results cover a much wider range of phase angles, from 6.9° to 87.6°. Hence our obtained slope parameter *G* should be more reliable.

## 4. Summary

The photometric observations of (25143) Itokawa were conducted from 2001 March to 2004 September. The absolute magnitude of this asteroid in Rband was determined to be $H_R$ = 19.09 ± 0.37 and its slope parameter is $G_R$ = 0.25 ± 0.29. These results are compatible with the known taxomonic type from spectral observation, S(IV)-type. By examining the lightcurve, the rotational period was found to be 12.1324±0.0001 hours, which is more accurate than the result derived only from observational data during the 2001 apparition (12.132 ± 0.0005 hours).

The asteroid (25143) Itokawa will make another apparition in 2006, while the spacecraft HAYABUSA will arrive there in September 2005, and perform in situ observations. The various scientific instruments might provide valuable information about Itokawa's detailed shape, surface roughness, and mineralogical composition, characteristics that have an effect on the lightcurve.